\documentstyle[12pt]{article}
\begin{document}

\title
{ Quark and gluon jet spectra \\
  in $\gamma\gamma$ collisions
        }
\author
{M.N.Dubinin\\
\it Institute for Nuclear Physics, Moscow State University\\
\it 119899 Moscow, Russia}
\date{}
\maketitle
\begin{abstract}
We consider three jet production at high $p_T$ in the
$\gamma\gamma \rightarrow
q \bar q g$ process and calculate quark and gluon jet spectra.
The possibility of quark-gluon jet separation is discussed and
compared with the $e^+ e^- \rightarrow q \bar q g$ case.
\end{abstract}

\section{Introduction}

Three jet production in $e^+ e^-$ annihilation was investigated in
details at PETRA and later at TRISTAN energies \cite{general} in connection
with the critical tests of
perturbative QCD. In particular the
difference between energy spectra of quark and gluon jets and quark
and gluon jet multiplicities provided important information for
quantitative tests of quantum chromodynamics.

Besides the studies of hadronic final state as a result of direct
$e^+e^-$ annihilation, the measurement of the two photon processes
$e^+e^- \rightarrow e^+e^-X$ at PETRA and TRISTAN shows high
physics potential. New tests are possible in LEP experiments,
especially the precise meauserement of the photon structure function
and hadron production at high $p_T$ \cite{LEP}

At present time new experimental possibilities connected with the
generation of $\gamma$ beams by Compton backscattering of the laser
beam on a high energy electron beam \cite{Ginzburg1} are
discussed extensively \cite{ee500}. Investigation of the jet phenomena in
$\gamma\gamma$
collisions at the energy of several hundred GeV is one of the most
interesting questions for $\gamma\gamma$ colliders phenomenology.

In this paper we discuss three jet production in the region of large
$p_T$ originated from t-channel quark exchange between the colliding
photons. At next linear colliders energy there are two other possible
mechanisms for
the jet production in high $p_T$ region with the cross sections
of the same order \cite{Ginzburg2}: (1) gluon exchange between
two pairs of quark jets (2) $W^+ W^-$ boson production. However,
they have different final state topologies and can easily be
distinguished from the mechanism under consideration. At smaller
$p_T$ the photon looks more like the hadron-like object and special
treatment beyond the simple QCD tree approximation is necessary.

The aims of this paper are the calculation of high $p_T$ quark
and gluon jets energy
spectra in $\gamma\gamma$ collisions and
comparison of the results with the $e^+ e^-$ case.

\section{Quark and gluon jet spectra in the process
         $\gamma\gamma \rightarrow q \bar q g$}

The first order QCD calculation for three jet production in $e^+ e^-
\rightarrow q \bar q g$ ($q=u,d,s$) gives the following well-known result
for the spectra \cite{ee}:
\begin{equation}
\frac{d\sigma}{dz_1dz_2}=\frac{16\pi\alpha^2\alpha_s}{9}
                         \frac{1}{s}\frac{z_1^2+z_2^2}{(1-z_1)(1-z_2)}
\end{equation}
where $z_1, z_2, z_3$ are the energy fractions of quark, antiquark
and gluon
\begin{equation}
z_1=\frac{2E_{\bar q}}{\sqrt{s}}, \hspace{5mm} z_2=\frac{2E_{q}}{\sqrt{s}},
             \hspace{5mm} z_3=\frac{2E_g}{\sqrt{s}}
\end{equation}
satisfying the equality
\begin{equation}
z_1+z_2+z_3=2
\end{equation}

Distribution (1) defines the topology of the final state (or topology
of the primary process) in $e^+ e^-$ collision. Since quarks and gluons are
unobservable, in the following various fragmentation models are
used to generate real event samples. However, event generation will be
beyond our analysis at present stage.

The calculation in $e^+e^-$ case for two Feynman amplitudes only (gluon
is radiated from quark or antiquark leg) is not technically difficult.
In $\gamma\gamma$ case we have 6 amplitudes represented in
Fig.1. Symbolic calculation for the corresponding 21 squared amplitudes
could be
more complicated. However, it is much simplified because the compact form
of the result for the sum of 21 squared diagrams can be
obtained \cite{compact}:
\begin{equation}
|M|^2=\frac{64}{27} e^4 {g_{ s}}^2 \,(p_4 p_5) \,
    \frac{ \sum_{i=1}^{3} \, (p_ip_4)( p_ip_5) [(p_ip_4)^2+(p_ip_5)^2]}
            {\prod_{i=1}^{3}\,(p_ip_4)\,(p_ip_5)}
\end{equation}
We reproduced this formula using CompHEP package \cite{CompHEP} for
generation of symbolic result and REDUCE system \cite{REDUCE} in
the following algebraic transformations.

For the process $\gamma(p_1)\gamma(p_2) \rightarrow
q(p_4) \bar q(p_5) g(p_3)$ we are using the kinematical variables

\begin{equation}
\quad {\rm cos} \vartheta_{14}^*, \quad z_1, \quad z_2, \quad \lambda
\end{equation}
where ${\rm cos} \vartheta_{14}^*$ is the angle between particle 1 and
particle 4 in the c.m.s. of (1,2), $\lambda$ is the helicity angle
between the planes (1,4) and (3,4) in the c.m.s of (3,5). The phase
space in these variables takes the form

\begin{equation}
dR_3=\frac{1}{1024\pi^4}  dz_1 dz_2 d{\rm cos}\vartheta_{14}^* d\lambda
\end{equation}

In the case of $e^+e^- \rightarrow q \bar q g$ process the matrix
element has collinear singuliarities only. In the case of
$\gamma(p_1)\gamma(p_2) \rightarrow q(p_4)\bar q(p_5) g(p_3)$ process
additional t-channel singuliarity appears.
The denominator of the squared amplitude contains four momenta products
(see more details in the Appendix)

\begin{eqnarray}
p_1p_4=z_2 \frac{s}{4} (1-{\rm cos}\vartheta^*) \\
p_1p_3=\frac{s}{4z_2}(a_1-b_1{\rm cos}\lambda)\\
p_2p_5=\frac{s}{4z_2}(a_4-b_1{\rm cos}\lambda)
\end{eqnarray}

where

\begin{eqnarray}
&&a_1={\rm cos}\vartheta^*(z_2^2+z_1z_2-2z_1-2z_2+2)+z_2(2-z_1-z_2) \\
&&a_4=-{\rm cos}\vartheta^*(-z_1z_2+2z_1+2z_2-2)+z_1z_2\\
&&b_1=2{\rm sin}\vartheta^*\sqrt{(1-z_1)(1-z_2)(z_1+z_2-1)}
\end{eqnarray}

Integration over $\lambda$ leads to the structures $\sqrt{a_1^2-b_1^2}$
and $\sqrt{a_4^2-b_1^2}$ in the denominator (see more details in
Appendix).
For this reason besides the poles of matrix element for the
scattering in forward-backward directions (no gluon emission
from final quark) there are parametrically
dependent poles at

\begin{equation}
|{\rm cos}\vartheta^*|=\frac{2(z_1+z_2-1)-z_1z_2}{z_1z_2}
\end{equation}
corresponding to the t-channel singuliarity in the diagrams with
gluon emission from the quark leg.
 (In this case the antiquark is emitted in the opposite hemisphere
at zero angle).
While integrating over ${\rm cos}\vartheta^*$ we introduced kinematical
cuts $\epsilon_1$ and $\epsilon_2$ for quark near the forward pole
and the poles defined by (13). After two integrations over the angular
variales $cos\vartheta^*,\,\lambda$ we obtain the following result
for three jet spectra in $\gamma\gamma$ collisions:

\begin{eqnarray}
&&\frac{d\sigma}{dz_1dz_2}=\frac{ e^4 g^2}{27\pi^2} \frac{1}{s}
    \frac{1}{z_1^2 z_2^2 (1-z_1)(1-z_2)}
  \{z_1^2 z_2^2 [(1-z_1)^2+(1-z_2)^2] \nonumber \\
&& \hspace{15mm}  \times [{\rm ln} \frac{1}{2\epsilon_1\epsilon_2^2}+2
\,{\rm ln}
              \frac{\omega_1\omega_2}{ z_1 z_2}
    - \frac{2(z_1+z_2-1)-z_1z_2}{z_1z_2}{\rm ln} \frac{\omega_1}{\omega_2}]
                                                      \nonumber \\
&& \hspace{15mm}    +   z_1^2 ((1-z_1)^2(1-z_2)^2+z_2^4)
                     {\rm ln} \frac{2}{\epsilon_1}         \nonumber \\
&& \hspace{15mm}        +z_2^2 ((1-z_1)^2(1-z_2)^2+z_1^4)
                     {\rm ln} \frac{\omega_1 \omega_2}{\epsilon_2^2}
\nonumber \\ && \hspace{15mm} +2z_1^2z_2^2(z_1^2+z_2^2)
                     {\rm ln} \frac{\omega_1}{2 z_1 z_2}
\nonumber \\
&& \hspace{15mm} +z_1^4z_2+z_1z_2^4 -5z_1^3z_2-5z_1z_2^3+4z_1^2z_2+4z_1z_2^2
                 +z_1^2z_2^3+z_1^3z_2^2 \nonumber \\
&& \hspace{15mm} +4z_1^3+4z_2^3-4z_1^2z_2^2 -2z_1^2-2z_2^2-2z_1^4-2z_2^4
                                                           \}\nonumber \\
&&
\end{eqnarray}
where
\begin{eqnarray}
&\omega_1=2(z_1+z_2-1)=2(1-z_3)& \\
&\omega_2=2(1-z_1)(1-z_2)&
\end{eqnarray}

Besides the collinear singuliarities in $\gamma\gamma$ case the spectrum
(14) has the factor $(z_1 z_2)$ in denominator and logarithmic terms
originating from t-channel quark exchange. Kinematical cut for
$\epsilon_1$ in the case of nonzero fermion masses is $2m_q/\sqrt{s}$,
it is easy to show that the cut for $\epsilon_2$ is of the same order.

\section{Quark and gluon jet separation}

We show double differential spectra of quark and gluon jets
in $e^+ e^-$ and $\gamma\gamma$ collisions
((1),(14)) at the energy scale $\sqrt{s} \sim 10^2 GeV$ in
Fig.2. In the $\gamma\gamma$ case the cross section is an order of
magnitude larger than for $e^+e^-$ case. The shape of the spectra
are similar.

In the case when there are no additional methods of quark and gluon
jet identification the spectra are not measurable separately and the
usual way of analysis is jet ordering in energy. In Fig.3 we show
single differential spectra of quark and gluon jets in the case of
$e^+e^-$ and $\gamma\gamma$ collisions. In the $e^+e^-$ case quark
and gluon jet spectra show large difference giving the possibility
of good jet discrimination. For instance, about 70\% of the jets at
$z$ less than $1/3$ are gluon jets and about 50\% of the jets at $z$
between $0.6$ and $0.9$ are quark jets. In the $\gamma\gamma$ case
the difference between the spectra is also well pronounced and
identification possibilities of the quark and qluon jets by energy
ordering seem not worse than for $e^+e^-$ case.
Improvement of jet identification in some experimental situations
could be provided by flavor tagging \cite{tag} or comparison of
$q \bar q g$ and $q \bar q \gamma$ event topologies [1].
Probably some combination of different methods can give optimal
results.
\begin{center}
{\bf Acknowledgements}
\end{center}
The author is grateful to M.Fontannaz, J.Fujimoto, J.P.Guillet,
K.Kato, T.Munehiza and Y.Shimizu for useful comments and discussions.
The work was partially supported by ISF (grant M9B000) and INTAS (grant
93-1180, contract 1010-CT93-0024).

\section*{Appendix}

In this section we shall show some details of the calculation
taking squared amplitudes (1) and (5) (see Fig.1) as an example.

Four momenta products in the variables (5) have the form

\begin{eqnarray*}
&&p_1p_2=\frac{s}{2}\\
&&p_3p_4=\frac{s}{2}(1-z_1)\\
&&p_3p_5=\frac{s}{2}(1-z_2)\\
&&p_4p_5=\frac{s}{2}(z_1+z_2-1)\\
&&p_1p_4=z_2 \frac{s}{4} (1-{\rm cos}\vartheta^*) \\
&&p_2p_4=z_2 \frac{s}{4} (1+{\rm cos}\vartheta^*) \\
&&p_1p_3=\frac{s}{4z_2}(a_1-b_1{\rm cos}\lambda)\\
&&p_1p_5=\frac{s}{4z_2}(a_2+b_1{\rm cos}\lambda)\\
&&p_2p_3=\frac{s}{4z_2}(a_3+b_1{\rm cos}\lambda)\\
&&p_2p_5=\frac{s}{4z_2}(a_4-b_1{\rm cos}\lambda)
\end{eqnarray*}

where

\begin{eqnarray*}
&&a_1={\rm cos}\vartheta^*(z_2^2+z_1z_2-2z_1-2z_2+2)+z_2(2-z_1-z_2) \\
&&a_2={\rm cos}\vartheta^*(-z_1z_2+2z_1+2z_2-2)+z_1z_2              \\
&&a_3=-{\rm cos}\vartheta^*(z_2^2+z_1z_2-2z_1-2z_2+2)+z_2(2-z_1-z_2) \\
&&a_4=-{\rm cos}\vartheta^*(-z_1z_2+2z_1+2z_2-2)+z_1z_2              \\
&&b_1=2{\rm sin}\vartheta^*\sqrt{(1-z_1)(1-z_2)(z_1+z_2-1)}
\end{eqnarray*}

It is worth noticing that this set is not invariant under the
transposition $z_1 \leftrightarrow z_2$ (as long as we are using the
$\bar q g$ c.m.s. as the reference frame). The matrix element
has the symmetry $quark \leftrightarrow antiquark$, so this symmetry
must be restored in the final result for the spectra (at least in
the nondivergent terms, artificial cuts near the poles can break the
symmetry). Squared diagrams (1) and (5) sum

\begin{eqnarray*}
\frac{128}{81} e^4 g_s^2 (\frac{p_1p_3}{p_1p_5\,p_3p_4}
                         +\frac{p_2p_3}{p_2p_4\,p_3p_5})
\end{eqnarray*}
after the integration over the helicity angle $\lambda$ takes the
form

\begin{eqnarray*}
\frac{d\sigma}{dz_1dz_2 cos\vartheta^*}&=&
\frac{256}{81s} e^4 g_s^2
\frac{-2(a_1+a_2)\sqrt{a_2^2-b_1^2}{\rm
arctg}(\sqrt{\frac{a_2-b_1}{a_2+b_1}}
                            {\rm tg}(\lambda/2)) +\lambda(a_2^2-b_1^2)}
     {(a_2^2-b_1^2)(z_1-1)}
\end{eqnarray*}
where

\begin{eqnarray*}
\sqrt{a_2^2-b_1^2}=2 | z_1z_2({\rm cos}\vartheta^*-1)+2(z_1+z_2-1) |
\end{eqnarray*}

Integration over ${\rm cos}\vartheta^*$ must take into account two cases for
the sign of the absolute value. The physical region of the reaction
$\gamma\gamma \rightarrow q \bar q g$ in $z_1,z_2$ plane is the
triangle with vertices (0,1);(1,0);(1,1).
Hyperbola
reflecting the relation between $cos\vartheta^*, z_1, z_2$ for t-channel
pole (13) crosses the physical region from (0,1) to (1,0) (see Berends
et.al in [7]).
Integration of the rational function

\begin{eqnarray*}
\frac{d\sigma}{dz_1 dz_2 dcos\vartheta^*}=\frac{128\pi e^4 g_s^2}{81s} \{
2\frac{2z_2-z_2^2+z_2^2{\rm cos}\vartheta^*}{ |z_1z_2(1-{\rm cos}\vartheta^*)
                    -2(z_1+z_2-1) |(1-z_1)}\\
-\frac{2}{1-z_1} +\frac{{\rm cos}\vartheta^*(z_2^2+z_1z_2-2z_1-2z_2+2)+
                         z_2(2-z_1-z_2)}
      {(1-{\rm cos}\vartheta^*)(1-z_2)z_2^2}\}
\end{eqnarray*}
gives the result for the spectrum

\begin{eqnarray*}
&&\frac{d\sigma}{dz_1dz_2}=\frac{e^4 g_s^2}{81\pi^2}\frac{1}{s}
\frac{1}{z_1^2 z_2^2 (1-z_1)(1-z_2)} \\
&&[ z_2^2 (1-z_2)^2 {\rm ln}(\frac{\omega_1 \omega_2}{\epsilon_2^2})
+z_1^2 (1-z_1)^2 {\rm ln}(\frac{2}{\epsilon_1}) \\
&&+z_1^2 z_2^2 (z_1+z_2-1)+(z_1^2 (1-z_1)+z_2^2 (1-z_2))(2z_1+2z_2-2-z_1z_2)]
\end{eqnarray*}

Calculation for the sum of 21 squared diagrams gives the same symbolic
structures.

%\documentstyle[epsf]{article}
%\begin{document}
%\newpage
%\eject
\vspace{20mm}
\section*{Figure captions}
\begin{itemize}
\item[Fig.1] {Feynman diagrams for $\gamma\gamma \rightarrow q \bar q g$.}

\item[Fig.2] {Double differential spectra $s\,d\sigma/dz_1dz_2$ at fixed
          $z_2=0.9$ of parent quark and gluon in
         the reactions $e^+e^- \rightarrow q \bar q g$ and
                       $\gamma\gamma \rightarrow q \bar q g$ ($q=u$).
       $\epsilon_1=\epsilon_2=10^{-5},\,e=0.313,\,g_s=1.59$ }

\item[Fig.3] {Differential spectra $s\,d\sigma/dz_2$ (integrated over $z_1$)
                     of parent quark and gluon in
                 the reactions $e^+e^- \rightarrow q \bar q g$
                  and $\gamma\gamma \rightarrow q \bar q g$ ($q=u$). }
\end{itemize}

\begin{thebibliography}{99}
\bibitem{general}
Collected physics papers (I) of TRISTAN experiments, KEK, Tsukuba, 1994
%Y.K.Kim et.al. (AMY collaboration), Phys.Rev.Lett.,
%63 (1989) 1772\\
%H.Takaki et.al. (VENUS collaboration), Phys.Rev.Lett., 71 (1993) 38
\bibitem{LEP}
P.\,Aurenche, A.\,Douiri, R.\,Baier, M.\,Fontannaz, D.\,Schiff, in:{\it
Physics at
LEP}, ed.by J.Ellis,R.Peccei, CERN report 86-02, 1986 \bibitem{Ginzburg1}
%I.Ginzburg, G.Kotkin, V.Serbo, V.Telnov, Pisma ZhETF 34 (1981) 514;
%JETP Lett.34 (1982) 491\\
I.\,Ginzburg, G.\,Kotkin, V.\,Serbo, V.\,Telnov, Nucl.Instr.Meth. 205
(1983) 147
\bibitem{ee500}
E.\,Boos, M.\,Dubinin, V.\,Ilyin, A.\,Pukhov, in:{\it $e^+e^-$
collisions at 500 GeV: the
physics potential}, ed.by P.Zerwas, DESY report 93-123C, 1993, p.561
\bibitem{Ginzburg2}
I.\,Ginzburg, D.\,Ivanov, V.\,Serbo, in:{\it Proc.of Workshop on Physics and
Experiments
with Linear $e^+e^-$ Colliders}, ed.by F.\,Harris, S.\,Olsen, S.\,Pakvasa,
X.Tata, World Scientific, Singapore, 1993, p.600
\bibitem{ee}
J.\,Ellis, M.K.\,Gaillard, G.G.\,Ross, Nucl.Phys.B111 (1976) 253\\
T.A.\,DeGrand, Y.J.\,Ng, S.H.H.\,Tye, Phys.Rev.D16 (1977) 3251\\
A.\,DeRujula, J.\,Ellis, E.G.\,Floratos, M.K.\,Gaillard, Nucl.Phys.B138
(1978) 387\\
G.\,Kramer, G.\,Schierholz, Phys.Lett.82B (1979) 102\\
P.\,Hoyer, P.\,Osland, H.G.\,Sander, T.F.\,Walsh, P.M.\,Zerwas, Nucl. Phys.
B161 (1979) 349\\
S.\,Nandi, W.\,Wada, Phys.Rev.D21 (1980) 76\\
H.P.\,Nilles, K.H.\,Streng, Phys.Rev.D23 (1981) 1944
\bibitem{compact}
F.\,Berends, Z.\,Kunszt, G.\,Gastmans, Phys.Lett.92B (1980) 186,
                                 Nucl.Phys.B182 (1981) 397\\
P.\,Aurenche, A.\,Douiri, R.\,Baier, M.\,Fontannaz, D.\,Schiff, Z.Phys.C24
(1984) 309
\bibitem{CompHEP}
E.\,Boos et.al.,in:{\it '91 Electroweak Interactions and Unified Theories
(Proc.
of the XXVIth Recontre de Moriond)}, ed.by J.\,Tran Than Van, Editions
Frontieres, 1991, p.501\\
E.\,Boos et.al.,in: {\it New Computing Techniques in Physics Research II
(Proc. of the Second Int.Workshop on Software Engineering, Artificial
Intelligence and Expert Systems in High Energy and Nuclear Physics)},
ed.by D.\,Perret-Gallix, World Scientific, 1992, p.665\\
E.\,Boos, M.\,Dubinin, V.\,Ilyin, A.\,Pukhov, V.\,Savrin,
preprint INP MSU 94-36/358, 1994 ({\bf hep-ph/9503280})
\bibitem{REDUCE}
REDUCE by A.C.\,Hearn, RAND Corp., CP78 (rev.7.91), 1991
\bibitem{tag}
H.\,Borner, P.\,Grosse-Wiesmann, in: {\it $e^+ e^-$ collisions at
500 GeV: the physics potential}, ed.by P.~Zerwas, DESY report 92-123A,
1992, p.63
\end{thebibliography}
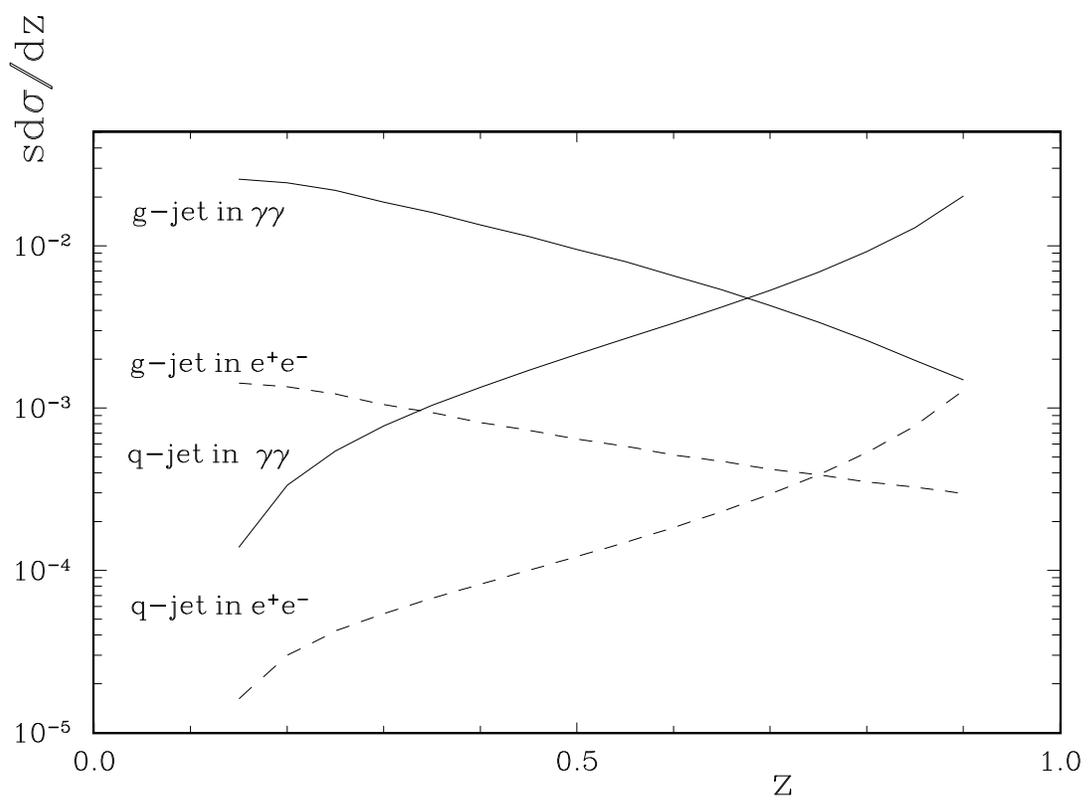
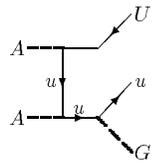
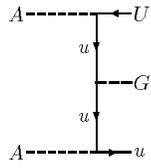
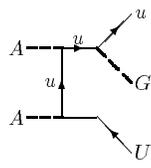
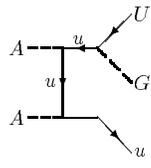
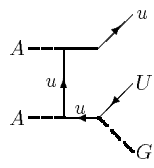
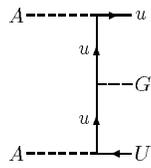
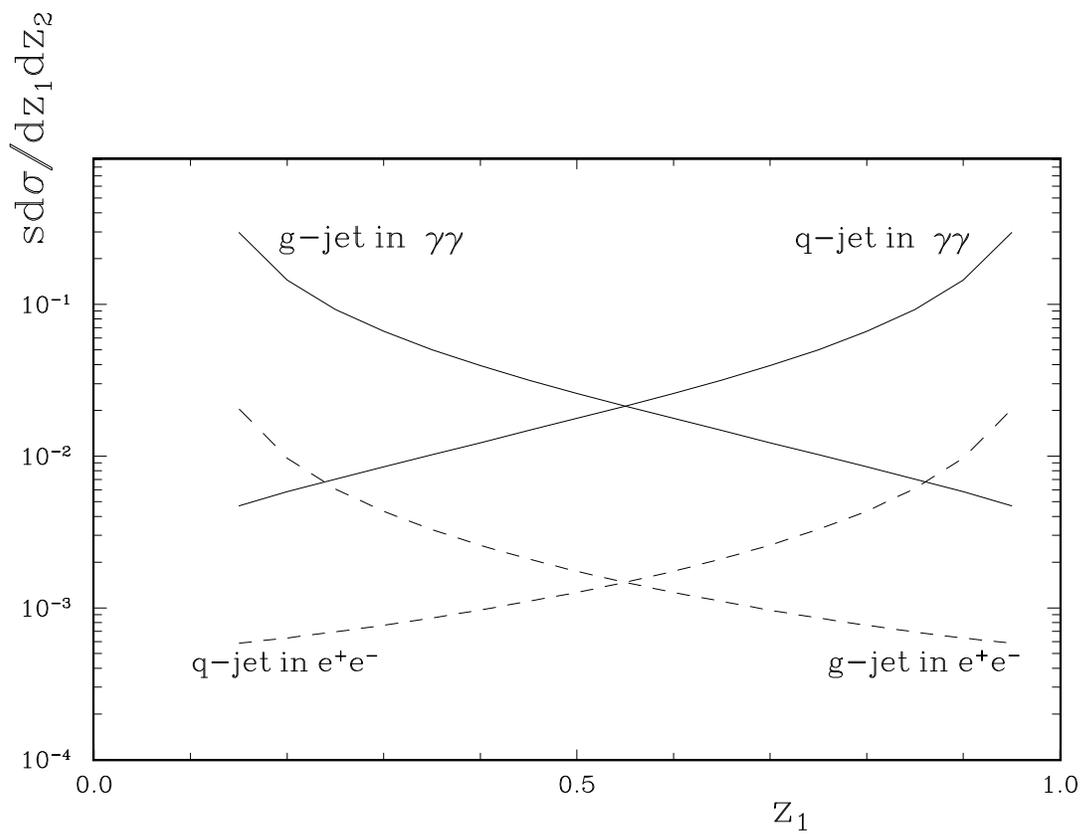
\end{document}